\documentclass[pra,aps,superscriptaddress,showpacs,twocolumn,amssymb,amsmath]{revtex4}
\usepackage{graphicx}
\usepackage{theorem}
\usepackage{dcolumn}
\usepackage{bm}
\usepackage{mathrsfs}
\usepackage{setspace}

\begin{document}

\title{Emergence of spiral dark solitons in the merging of rotating Bose-Einstein condensates}

\author{Toshiaki Kanai}
\affiliation{Department of Physics, Osaka City University, 3-3-138 Sugimoto, Sumiyoshi-Ku, Osaka 558-8585, Japan}

\author{Wei Guo}
\email[Email: ]{wguo@magnet.fsu.edu}
\affiliation{National High Magnetic Field Laboratory, 1800 East Paul Dirac Drive, Tallahassee, FL 32310, USA}
\affiliation{Mechanical Engineering Department, Florida State University, Tallahassee, FL 32310, USA}

\author{Makoto Tsubota}
\affiliation{Department of Physics, Osaka City University, 3-3-138 Sugimoto, Sumiyoshi-Ku, Osaka 558-8585, Japan}
\affiliation{The OCU Advanced Research Institute for Natural Science and Technology (OCARINA), Osaka City University, 3-3-138 Sugimoto, Sumiyoshi-Ku, Osaka 558-8585, Japan}

\date{\today}

\begin{abstract}
Merging of isolated Bose-Einstein condensates (BECs) is an important topic due to its relevance to matter-wave interferometry and the Kibble-Zurek mechanism. Many past research focused on merging of BECs with uniform initial phases. In our recent numerical study (Phys. Rev. \textbf{A 97}, 013612 (2018)), we revealed that upon merging of rotating BECs with non-uniform initial phases, spiral-shaped dark solitons can emerge. These solitons facilitate angular momentum transfer and allow the merged condensate to rotate even in the absence of quantized vortices. More strikingly, the sharp endpoints of these spiral solitons can induce rotational motion in the BECs like vortices but with effectively a fraction of a quantized circulation. This paper reports our systematic study on the merging dynamics of rotating BECs. We discuss how the potential barrier that initially separates the BECs can affect the profile of the spiral solitons. We also show that the number of spiral solitons created in the BECs matches the relative winding number of the rotating BECs. The underlying mechanism of the observed soliton dynamics is explained.
\end{abstract}
 \maketitle

\section{Introduction}
Merging of isolated Bose-Einstein condensates (BECs) has been a topic of extensive experimental and theoretical studies. The motivations of these studies include better understanding of the processes involved in matter wave interferometry \cite{Andrews-1997-Sci,Shin-2004-PRL, Hadzibabic-2004-PRL, Shin-2005-PRA}, investigating interesting nonlinear quantum hydrodynamics \cite{Liu-2000-PRL, Yang-2013-PRA}, and exploring the creation of topological phase defects (i.e., quantized vortices) \cite{Stock-2005-PRL, Scherer-2007-PRL, Xiong-2013-PRA} and nontopological phases defects (i.e., dark solitons) \cite{Yang-2007-PRA, Yang-2008-PRA,Toikka-2014-JPB}. Especially, the creation of phase defects upon merging of BECs has been utilized to test the celebrated Kibble-Zurek mechanism \cite{Weiler-2008-Nature, Carretero-2008-PRA, Corman-2014-PRL,Lamporesi-2013-Nat}. This mechanism explains that the formation of phase defects following a rapid second-order phase transition is due to the merging of isolated superfluid domains with random relative phases \cite{Zurek-1996-PR, Kibble-2007-PT}. So far, many studies on condensate merging have focused on cases with condensates having uniform initial phases. In recent numerical work, we studied the merging of two concentric condensates with axial symmetry in two-dimensional (2D) space. One of the two condensates was set to carry a finite angular momentum with non-uniform initial phases before merging occurs. We were interested in how the angular momentum could be transferred from the rotating condensate to the initially static condensate and whether this transfer was accompanied by the creation and transfer of quantized vortices. Some highlights of the simulation results are briefly reported in a recent paper~\cite{Kanai-2018-PRA}. We observed the emergence of a spiral dark soliton with a sharp inner endpoint during the merging process. This spiral dark soliton enables the transfer of angular momentum between the two concentric condensates and allows the merged condensates to rotate even in the absence of quantized vortices. More strikingly, an examination of the flow field around the sharp endpoint of the dark soliton reveals that this endpoint can induce rotational motion in the condensate like a vortex point, but with effectively a fraction of a quantized circulation that matches the phase step across the soliton boundary. In order to understand this intriguing soliton dynamic, a systematic study of the merging process of 2D concentric condensates under various conditions is indispensable. In this paper, we report such a detailed study.

In Sec. \ref{Sec-II}, we discuss the Gross-Pitaevskii equation (GPE) and the numerical model used in our simulation work. In Sec. \ref{Sec-III}, we present our simulation results. The effect of the potential barrier, which initially separates the two concentric BECs, on the merging dynamics is discussed. The interesting rotational motion around the sharp endpoints of the spiral solitons is re-visited. We then show how the number of spiral solitons created in the condensate is controlled by the initial relative winding number of the two condensates. In Sec. \ref{Sec-IV}, we discuss the underlying mechanism for the formation of the spiral solitons and also comment on how the merged condensate can possess angular momentum without quantized vortices. A brief summary is given in Sec. \ref{Sec-V}.

\section {Numerical method} \label{Sec-II}
The dynamical evolution of a BEC at zero temperature can be described accurately by a non-linear Gross-Pitaevskii equation (GPE) \cite{Pitaevskii-2003-book}. For BECs confined to two dimensional space, this equation is given by:
\begin{equation}
i\hbar \frac{\partial \psi}{\partial t}=\left[-\frac{\hbar^2}{2M}\nabla^2+V(\textbf{r},t)+g|\psi|^2\right]\psi,
\label{Eq1}
\end{equation}
where $\hbar$ is Planck's constant, $M$ is the mass of the particles that form the condensate, $\psi$ is the condensate wave function, $V$ is the external potential that confines the condensate, and $g$ is the coupling constant that measures the strength of the particle interactions. In numerical simulations, it is convenient to re-write the above GPE in a dimensionless form by introducing dimensionless parameters $r=\xi{\tilde{r}}$, $t=(\hbar/ng){\tilde{t}}$, and $\psi=(\sqrt{N}/\xi)\tilde{\psi}$, where $\xi=\hbar/\sqrt{2Mng}$ is the healing length, $N=\int dS|\psi|^2$ is the total number of particles, and $n=N/S$ is the particle number density averaged over the system area $S$. It is straightforward to show that the original GPE now takes the following form:
\begin{equation}
i\frac{\partial \tilde{\psi}}{\partial \tilde{t}}=\left[-\tilde{\nabla}^2+\tilde{V}(\tilde{\textbf{r}},\tilde{t})+\tilde{g}|\tilde{\psi}|^2\right]\tilde{\psi}.
\label{Eq2}
\end{equation}
The dimensionless coupling constant $\tilde{g}$ is given by $\tilde{g}=N/(n\xi^2)$ and simply equals the dimensionless BEC area $S/\xi^2$. The potential $\tilde{V}=V/ng$ now measures the ratio of the external potential $V$ to the particle interaction strength ($ng$).

\begin{figure}[htb]
\includegraphics[scale=1.33]{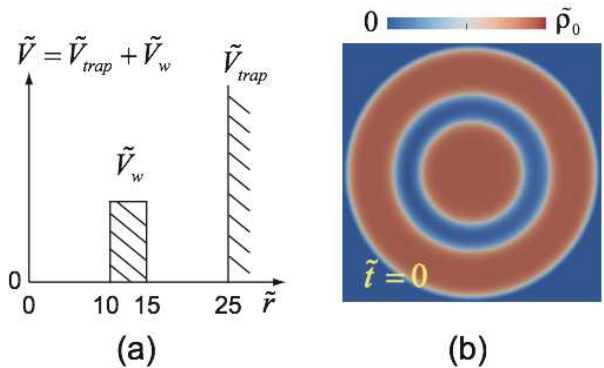}
\caption{(color online). (a) Schematic of the potential $\tilde{V}(\tilde{\textbf{r}},\tilde{t})$ used in our GPE simulation. (b) The initial density profile of the condensates when there is no rotation in the disk and the ring condensates.} \label{Fig1}
\end{figure}

Our goal is to study the merging process of a disk condensate with a concentric ring condensate. In our simulation, we set $\tilde{V}=\tilde{V}_{trap}+\tilde{V}_w$, where $\tilde{V}_{trap}$ represents the cylindrical hard-wall box potential that traps the condensates and $\tilde{V}_w$ denotes the potential barrier that separates the disk and the ring condensates, as shown in Fig.~\ref{Fig1} (a). The potential trap $\tilde{V}_{trap}$ has a radius of $25\xi$. This size is within the range of typical condensate sizes in real experiments (i.e., about $10\xi-10^2\xi$ \cite{Kwon-2016-PRL, Scherer-2007-PRL, Jendrzejewski-2014-PRL, Corman-2014-PRL}). The potential barrier $\tilde{V}_w$ has a square profile with a width of $5\xi$ and is located at a position such that the inner disk condensate is confined to have a radius of $10\xi$. Note that in real experiments, the height of the potential $V$ for trapping or separating the condensates depends on the frequency and the power flux of the laser beams, and the product $ng$ depends on the particle species and the particle number density. Using the parameters reported in literature, we find that the typical values for the dimensionless potential $\tilde{V}$ is in the range of $1-10^2$ \cite{Kwon-2016-PRL, Scherer-2007-PRL, Jendrzejewski-2014-PRL, Corman-2014-PRL, Killian-1998-PRL, Gaunt-2013-PRL}. In our simulation, we vary the height of $\tilde{V}_w$ from 1 to 10 in order to examine the barrier height effect on the condensate merging dynamics.

Our simulation is carried out in a region $\tilde{\textbf{r}}\in[-25,25]\times[-25,25]$ with a mesh grid of $500\times500$ nodes to ensure spatial convergence. The dimensionless time step $\delta_t$ is chosen to be $1.0\times10^{-4}$. To prepare the steady initial state, we first evolve Eq.~\ref{Eq2} in imaginary time \cite{Chiofalo-2000-PRE}. Fig.~\ref{Fig1} (b) shows an example of a typical initial condensate density profile when there is no rotation in the disk and the ring condensates. We can also set the condensates into rotation by either introducing a quantized vortex point at the center of the disk condensate or prescribing a phase gradient in the ring condensate such that the ring condensate carries a supercurrent. The circulation associated with the supercurrent can be any integer $m$ multiplied by the quantum circulation $\kappa=h/M$. At time $\tilde{t}=0$, we then suddenly remove the energy barrier $\tilde{V}_w$ and let the two condensates merge. The dynamical evolution of the condensate wavefunction during merging can be obtained by numerically integrating Eq.~\ref{Eq2} using an alternating direction implicit method \cite{Press-1992-book}.

We would like to emphasize that the condensate configurations adopted in our simulation can be easily realized in BEC experiments. For instance, Corman \emph{et al.} \cite{Corman-2014-PRL} and Eckel \emph{et al.} \cite{Eckel-2014-PRX} have utilized the interference patterns of a ring condensate and a disc condensate during free expansion to study the Kibble-Zurek mechanism and superfluid weak links. Their setup can be adapted to examine the results of our simulation.

\section {Merging of Concentric 2D Condensates}\label{Sec-III}
\subsection {Effect of Potential Barrier Height}
\begin{figure*}[htb]
\includegraphics[scale=1.4]{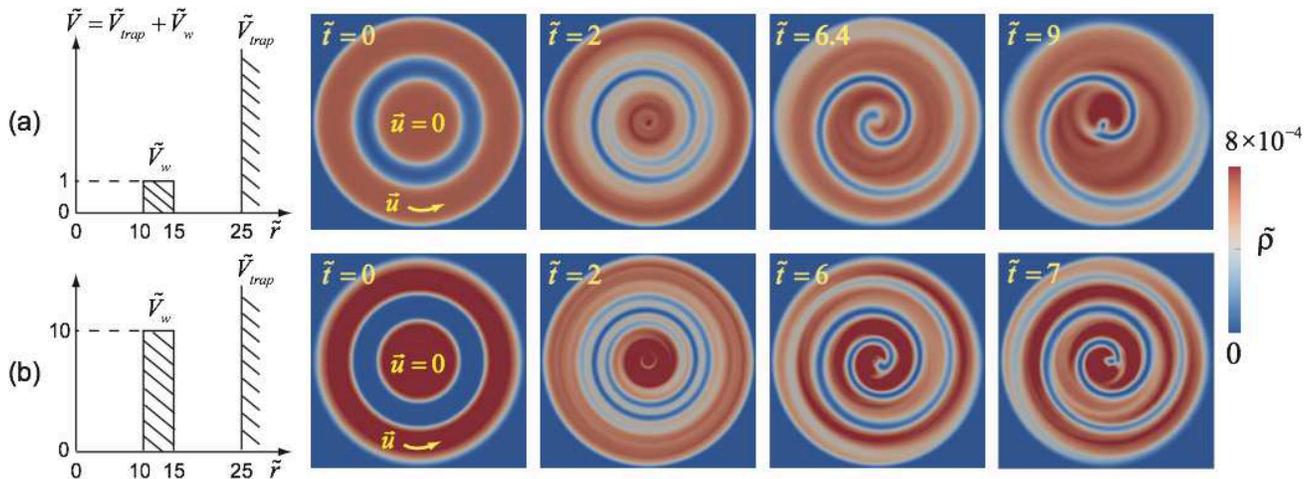}
\caption{(color online). Schematic of the potential $\tilde{V}$ and representative snapshots showing the time evolution of the BEC density $\tilde{\rho}$ for (a) $\tilde{V}_w=1$ and (b) $\tilde{V}_w=10$. The initial state is a static inner disk condensate with a rotating outer ring condensate carrying a circulation of $\kappa$.} \label{Fig2}
\end{figure*}
\begin{figure*}[htb]
\includegraphics[scale=1.4]{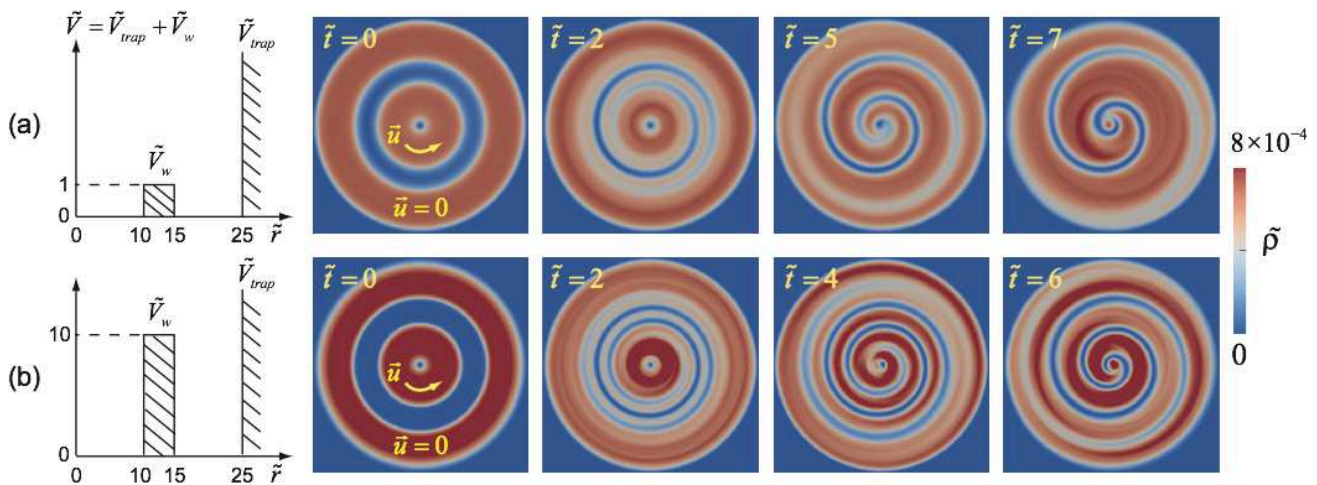}
\caption{(color online). Schematic of the potential $\tilde{V}$ and representative snapshots showing the time evolution of the BEC density $\tilde{\rho}$ for (a) $\tilde{V}_w=1$ and (b) $\tilde{V}_w=10$. The initial state is a static outer ring condensate with a rotating inner disk condensate having a single vortex point at the center.} \label{Fig3}
\end{figure*}
We first study the effect of the potential barrier height $\tilde{V}_w$ on the condensate merging dynamics. Fig.~\ref{Fig2} shows representative snapshots of the dynamical evolution of the dimensionless condensate density $\tilde{\rho}=|\tilde{\psi}|^2$, following the removal of the potential barrier $\tilde{V}_w$ for the case where initially the inner disk condensate is static and the outer ring condensate carries a supercurrent with a circulation of $\kappa$. Fig.~\ref{Fig2} (a) is for $\tilde{V}_w=1$ and Fig.~\ref{Fig2} (b) is for $\tilde{V}_w=10$. Similar to what we have reported~\cite{Kanai-2018-PRA}, a spiral stripe with depleted condensate density emerges in both cases. Across this stripe there is an abrupt phase step $\triangle\phi$. This stripe is a dark soliton that is similar in nature to the ring dark solitons identified in the expansion of 2D disk and annular condensates \cite{Yang-2007-PRA, Yang-2008-PRA, Toikka-2014-JPB}. It is known that a soliton stripe can drift in the condensate with a velocity $v_s$ that is controlled by the phase step $\triangle\phi$ as $v_s=v_0\cdot{\texttt{cos}(\triangle\phi/2)}$, where $v_0=(g\rho/M)^{1/2}$ is the Bogoliubov speed of sound \cite{Jackson-1998-PRA}. For a ``black'' soliton with complete density depletion (i.e., $\tilde{\rho}$=0 at the center), $\triangle\phi=\pi$ and the soliton has zero velocity with a thickness on the order of $\xi$. When $\triangle\phi$ decreases, the soliton becomes shallower and wider, and its speed increases. Comparing the soliton profiles in the snapshots for $\tilde{V}_w=1$ and $\tilde{V}_w=10$, one can see that in both cases the spiral soliton develops a sharp inner endpoint that spirals toward the condensate center. But the length of the soliton stripe for $\tilde{V}_w=10$ is obviously longer than that for $\tilde{V}_w=1$. This result is probably natural since in the $\tilde{V}_w=10$ case the initial condensate density at $\tilde{t}=0$ drops steeply in the potential barrier region. This large density gradient provides more potential energy for the formation of the soliton after $\tilde{V}_w$ is removed. As the sharp inner end of the soliton stripe approaches the center, snake instability occurs when the local curvature radius of the inner end becomes comparable to $\xi$ \cite{Mamaev-1996-PRL, Theocharis-2003-PRL, Ma-2010-PRA}, and a vortex point is nucleated near the center.

We have also studied the merging dynamics for an initially static outer ring condensate with a rotating inner disk condensate that has a single vortex point at the center. The results are shown in Fig.~\ref{Fig3} for both $\tilde{V}_w=1$ and $\tilde{V}_w=10$ cases. Again, a spiral soliton emerges and the length of the soliton stripe appears to be longer for $\tilde{V}_w=10$. As the inner end of the soliton stripe approaches the center, the vortex point can merge into the soliton stripe, which renders the condensate completely vortex free~\cite{Kanai-2018-PRA}, as depicted in Fig.~\ref{Fig3} (i.e., at $\tilde{t}=7$ for $\tilde{V}_w=1$ case and at $\tilde{t}=6$ for $\tilde{V}_w=10$ case). In long time evolution, the solitons observed in all cases eventually decay into vortices via snake instability.

\subsection {Transfer of Angular Momentum and Rotation}
\begin{figure}[htb]
\includegraphics[scale=1.24]{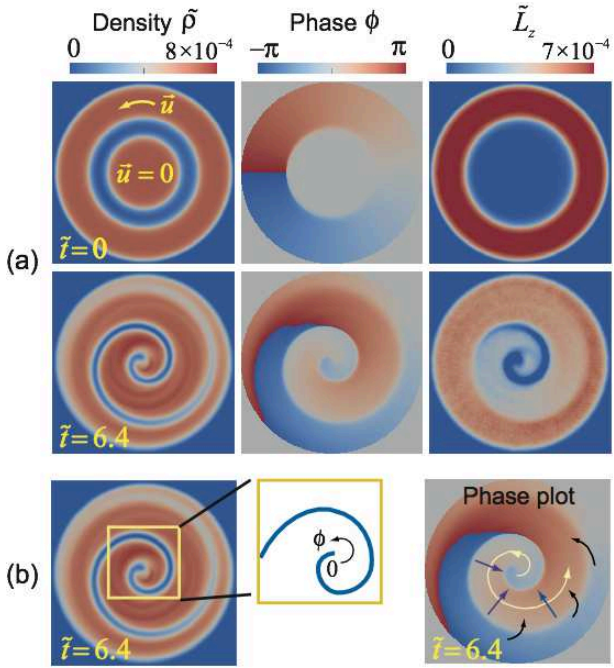}
\caption{(color online). (a) Snapshots of the condensate density, phase, and angular momentum density $\tilde{L}_z$ for the case shown in Fig.~\ref{Fig2} (a). (b) Schematics illustrating the underlying mechanism for the mass and angular momentum transfer.}
\label{Fig4}
\end{figure}

To study the angular momentum transfer, we define a dimensionless angular momentum density $\tilde{L}_z$ as
\begin{equation}
\tilde{L}_z=\frac{\xi^2}{\hbar N}\left(\psi^*\hat{L}_z\psi\right)=\frac{1}{i}\tilde{\psi}^*(\tilde{x}\frac{\partial}{\partial \tilde{y}}-\tilde{y}\frac{\partial}{\partial \tilde{x}})\tilde{\psi}.
\end{equation}
As reported in our previous publication~\cite{Kanai-2018-PRA}, the angular momentum initially contained in the rotating condensate can spread to the initially static condensate region along the spiral channel formed by the soliton stripe. Fig.~\ref{Fig4} (a) shows example snapshots of the condensate density $\tilde{\rho}$, phase $\phi$, and angular momentum density $\tilde{L}_z$ for the $\tilde{V}_w=1$ case reported in Fig.~\ref{Fig2} (a). The transfer of angular momentum can occur in the absence of quantized vortices while the total angular momentum is conserved. A careful examination of the flow shown in Fig.~\ref{Fig4} (a) reveals something interesting. The flow in the initially rotating condensate is counterclockwise so cannot enter the outward spiral channel formed by the soliton. Therefore, the rotational motion in the initially static disk condensate must be induced by a different mechanism that is effective even without quantized vortices.

\begin{figure}[htb]
\includegraphics[scale=1.29]{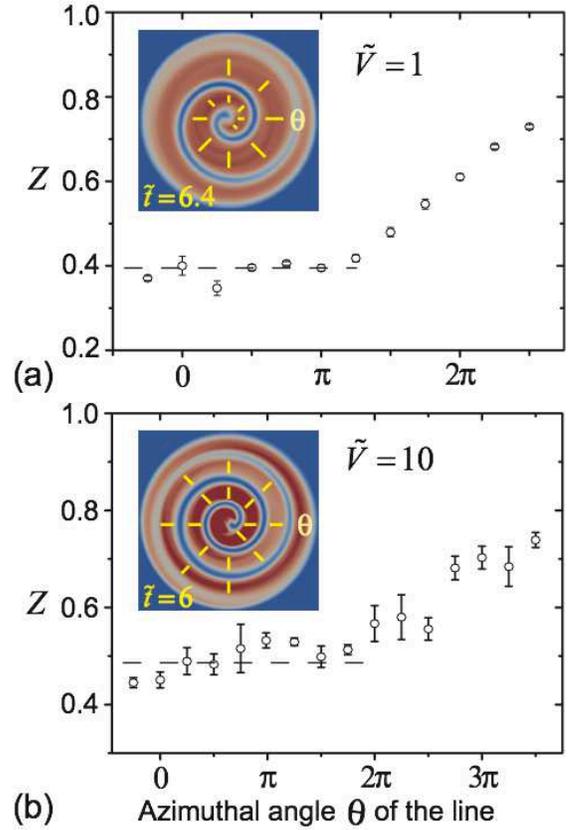}
\caption{(color online). Calculated vortex charge $Z$ in the condensate along the solid yellow lines that are shown in the inset for (a) $\tilde{V}_w=1$ and (b) $\tilde{V}_w=10$.} \label{Fig5}
\end{figure}

This novel mechanism has been identified as due to the induced flow by the sharp endpoint of the spiral soliton~\cite{Kanai-2018-PRA}. As illustrated in Fig.~\ref{Fig4} (b), due to the phase step across the boundary of the soliton, there exists a phase winding of $\triangle\phi$ around the sharp inner endpoint. This phase winding leads to a rotational motion in the condensate, making the sharp endpoint effectively a ``vortex point'' that carries a fraction of a quantized circulation given by $(\frac{\triangle\phi}{2\pi})\kappa$. We should emphasize that mathematically the circulation, i.e., integral of the velocity along a closed contour around the endpoint, is still zero due to the opposite phase velocity inside the soliton density depleted region. The flow induced by the endpoint carries the condensate mass from the inner region to the outer region guided by the spiral channel, which leads to a phase increment along the soliton boundary. Consequently, this phase increment leads to a radial phase gradient in the condensate that drives an inward mass flow. In the shallow tail part of the soliton stripe, the condensate density in the soliton is not depleted and mass flow from the outer region through the soliton boundary towards the inner region becomes significant. A mass circulation that effectively mixes the condensates and transports angular momentum is established.

\begin{figure*}[htb]
\includegraphics[scale=1.56]{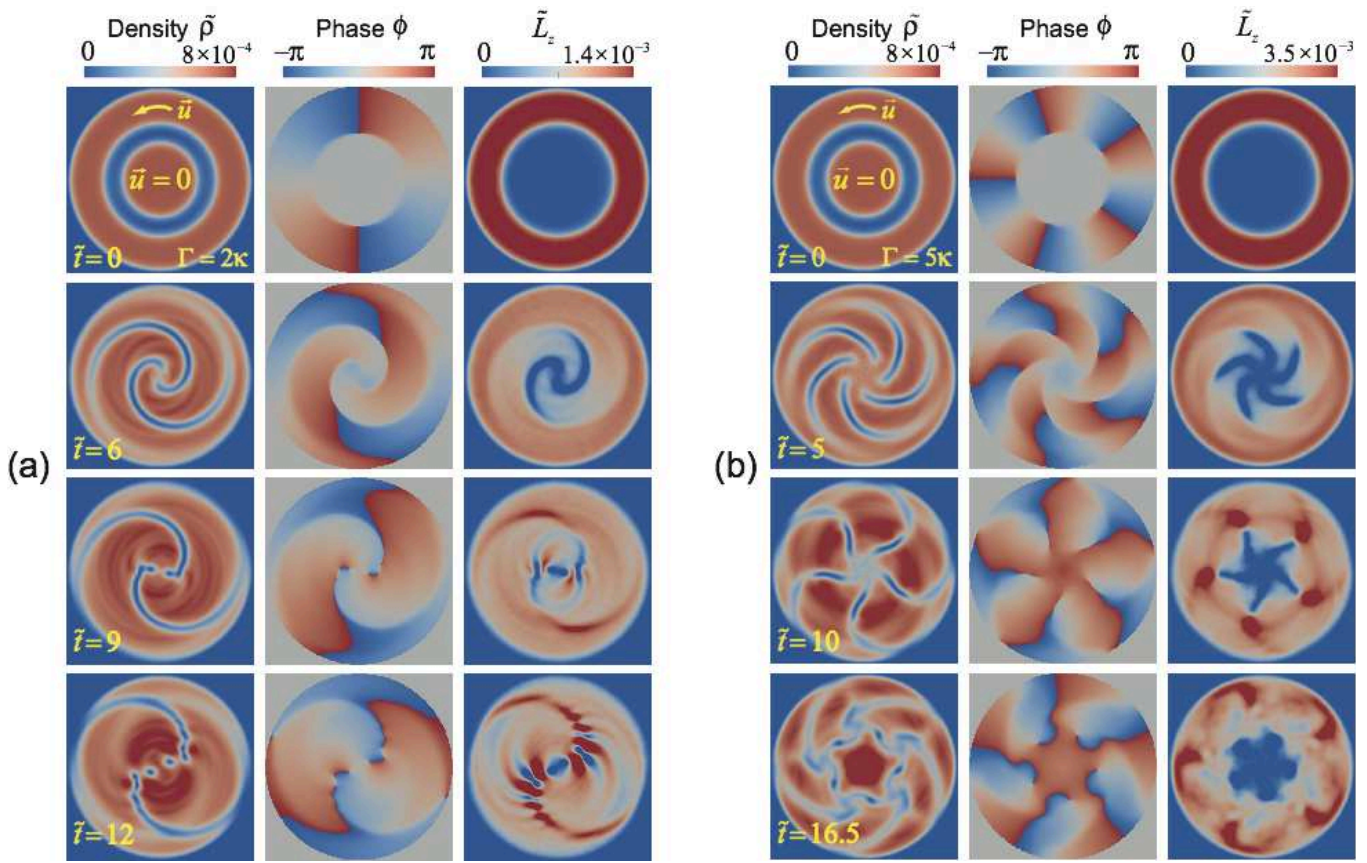}
\caption{(color online). Time evolution of the condensate density, phase, and angular momentum density $\tilde{L}_z$ when the static inner disk condensate merges with the rotating outer ring condensate that carries a circulation of (a) $2\kappa$ and (b) $5\kappa$. The potential barrier $\tilde{V}_w=1$.} \label{Fig6}
\end{figure*}

To show that the inner endpoint does behave like a vortex point with a fractional quantized circulation, we introduce a vortex charge parameter $Z$, defined as $Z=\frac{m}{\hbar}|\textbf{r}\times\textbf{v}(\textbf{r})|=\frac{m}{\hbar}rv_{\theta}$, where $v_{\theta}$ is the velocity along the azimuthal angle direction. For the flow field induced by a vortex point at the center, $Z$ is a constant and equals the winding number of the vortex. In Fig.~\ref{Fig5}, we show the calculated $Z$ values along some radial lines in the condensate when the inner endpoint of the spiral soliton is close to the center. Near the endpoint, the $Z$ values are about 0.4 for the $\tilde{V}_w=1$ case and about 0.48 for the $\tilde{V}_w=10$ case, which matches well with the measured phase step across the soliton boundary near the endpoint (i.e., $\triangle\phi\simeq0.8\pi$ for $\tilde{V}_w=1$ and $\triangle\phi\simeq0.95\pi$ for $\tilde{V}_w=10$). In the tail region of the soliton where there are appreciable mass flows across the soliton boundary from the outer region, $Z$ starts to increase towards one, a value expected for the flow in the initial ring condensate.

\subsection {Merging of Condensates with Multiple Quantum Circulations}
We have also examined the merging of a static inner disk condensate with a rotating outer ring condensate that carries a uniform (i.e. axially symmetric) supercurrent with multiple quantum circulations. Fig.~\ref{Fig6} shows typical snapshots of the time-evolution of $\tilde{\rho}$, $\phi$, and $\tilde{L}_z$ for the outer ring condensate having a circulation of $2\kappa$ (in Fig.~\ref{Fig6} (a)) and $5\kappa$ (in Fig.~\ref{Fig6} (b)) with $\tilde{V}_w=1$. Interestingly, multiple spiral dark solitons are observed in these cases. The number of soliton stripes match exactly the winding number of the flow (i.e., number of quantum circulations) in the initial ring condensate. These spiral solitons also develop sharp inner endpoints that induce rotational motion in the initially static disk condensate, aiding the transfer of angular momentum in a similar way as we have discussed in the previous section. As the local curvature radius near the inner ends of the solitons keeps reducing, vortex points peel off from the soliton inner ends. Note that when there are multiple soliton stripes and vortices in the condensate, the flow field becomes very complex. Each soliton stripe can be strongly disturbed by the flows induced by nearby solitons and vortices. Consequently, snake instability tends to occur more easily and the inner ends of the solitons can break up into segments and vortices (e.g., see Fig.~\ref{Fig6} (a) at $\tilde{t}=12$). \begin{figure*}[htb]
\includegraphics[scale=1.33]{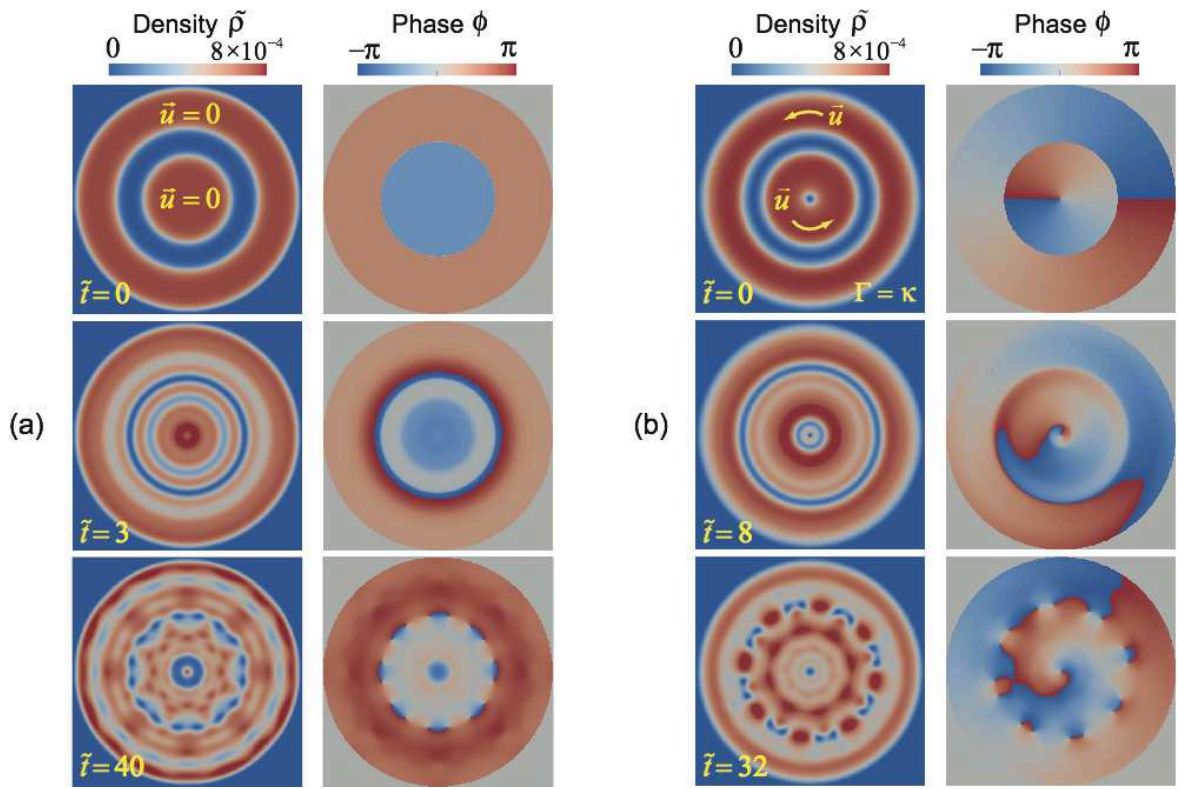}
\caption{(color online). Snapshots showing the time evolution of the condensate density and phase during merging when the initial phase difference between the ring and the disk condensates is constant across their interface. The potential barrier $\tilde{V}_w=1$. (a) is for the case where both condensates are initially static. (b) is for the case where the rotating ring condensate carries a circulation of $\kappa$ merging with a co-rotating disk condensate that has a single quantized vortex point at the center.} \label{Fig7}
\end{figure*}
In both cases presented in Fig.~\ref{Fig6}, the net vorticity (i.e., counting all positive and negative vortices) in the condensate always matches the number of quantum circulations in the initial outer ring condensate.

Further, we have tested cases where a static outer ring condensate merges with a rotating inner disk condensate that has multiple vortex points in it. Similarly, spiral solitons emerge from the condensate during merging. The number of soliton stripes again match the number of vortices in the initial disk condensate. It appears that the number of spiral soliton stripes created during condensate merging is controlled by the relative winding number (or relative circulation) of the two condensates. We shall discuss the underlying mechanism that is responsible for this interesting soliton formation process in Sec.~\ref{Sec-IV}.

\subsection {Merging of Condensates with Constant Phase Difference}
The observations presented in the previous section suggest that the number of spiral solitons is controlled by the relative winding number of the flows in the disk and the ring condensates. It is therefore natural to ask what can happen if the two condensates merge from an initial configuration that has no relative motion. One would expect that no spiral soliton should emerge in this situation. We have examined two representative cases as shown in Fig.~\ref{Fig7}. The first case (Fig.~\ref{Fig7} (a)) is for a static ring condensate merging with a static disk condensate. The second case (Fig.~\ref{Fig7} (b)) is for a rotating ring condensate that carries a circulation of $\kappa$ merging with a co-rotating disk condensate that has a single quantized vortex point at the center. In both cases, instead of spiral solitons, ring-shaped dark solitons are formed at the interface of the two condensates upon merging. These ring dark solitons expand in radius, bounce back from the trap boundary, and shrink toward the center of the condensate. They can undergo such expansion-shrinking cycles many times. Despite the apparent differences in geometry, this observed soliton dynamic is indeed very similar to the formation and propagation of planar solitons observed in the merging of 3D condensates with constant phase difference \cite{Shomroni-2009-NP}. At long time evolution, fluctuations in the simulation build up such that snake instability occurs. Eventually, the ring solitons in both cases break up into pairs of positive and negative vortex points (e.g., see Fig.~\ref{Fig7} (a) at $\tilde{t}=40$ and Fig.~\ref{Fig7} (b) at $\tilde{t}=32$).

A notable difference between the ring solitons and the spiral solitons reported in the previous sections is that the spiral solitons have sharp inner endpoints which can induce rotational motion in the condensate like vortices. These sharp endpoints make the angular momentum transfer between the two condensate regions possible. When the two condensates are static or co-rotate without any relative motion, angular momentum transfer between them is no longer necessary and the ring solitons remain intact instead of breaking up to develop the spiral shape. We therefore emphasize that although soliton formation upon merging of condensates is a well known phenomenon, the current simulation has confirmed an important point that it is the relative motion between the condensates prior to merging that causes the solitons to break and develop sharp endpoints. This novel soliton feature would not appear and be observed in early simulation works that focused on merging of condensates with uniform phases and no relative motion.

\section {Discussion}\label{Sec-IV}
\begin{figure}[htb]
\includegraphics[scale=1.24]{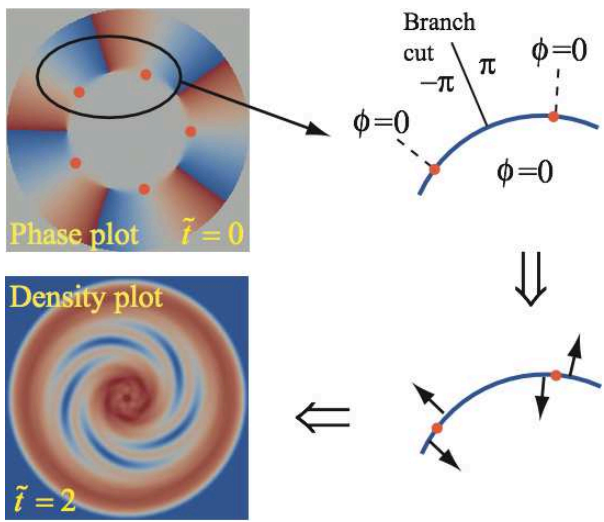}
\caption{(color online). Schematics showing the underlying mechanism on how a soliton at the interface of the two condensate domains breaks up and develops multiple spiral stripes.} \label{Fig8}
\end{figure}

The simulation results presented in the previous sections provide us some clues to understand why the spiral solitons can form with the total number of soliton stripes matching exactly the relative winding number of the two condensates. As an example, let us consider the phase plot at $\tilde{t}=0$ for the case where the outer ring condensate carries a supercurrent with a circulation of $5\kappa$, as shown in Fig.~\ref{Fig8}.
The initial phase of the inner disk condensate is constant (i.e., $\phi=0$) while the phase in the outer ring condensate increases counterclockwise between the branch-cut lines. The phase difference $\Delta{\phi}$ across the interface between the two condensates therefore varies along the interface. It is clear that there are five points (marked as red dots in the schematics) across which the phase difference $\Delta{\phi}$ changes sign. Note that in a condensate, a soliton stripe travels opposite to the direction of the phase step $\Delta{\phi}$ across the soliton boundary at a speed give by $v_s=v_0\cdot{\texttt{cos}(\triangle\phi/2)}$. Therefore, for a soliton stripe forming at the interface between the ring and the disk condensates, the stripe on either side of a red dot tends to move in opposite directions due to the change in sign of the phase step. Consequently, the soliton stripe must break up at these locations. Note that the breaking up of the soliton stripe occurs simultaneously with the formation of the soliton. Therefore, one would not see the formation of a complete ring soliton and then its breaking up into five pieces. Instead, one sees the gradual development of five soliton stripes. These soliton stripes have two ends, with one end spiraling in and the other end extending out.

\begin{figure}[htb]
\includegraphics[scale=1.33]{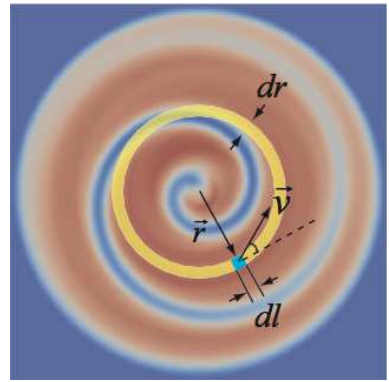}
\caption{(color online). Schematic for evaluating the angular momentum contained in the merged condensate in the axially symmetric trap.} \label{Fig9}
\end{figure}

In our simulation, we have observed that the merged condensate can carry angular momentum in the axially symmetric potential trap even in the absence of quantized vortices. This observation may appear a little counterintuitive at first sight. We would like to make some brief comments to clarify the relevant concepts. Let us consider the angular momentum carried by a ring element in the condensate, located between $r$ and $r+dr$, as shown schematically (i.e., the yellow ring area) in Fig.~\ref{Fig9}. The angular momentum possessed by this ring area can be calculated as:
\begin{equation}\label{Eq-AM}
dL_z=\int{\rho{dldr}\left[\hat{e}_z\cdot\left({\vec{r}\times\vec{v}}\right)\right]}=rdr\left[\oint{\rho{d\vec{l}}\cdot\vec{v}}\right]
\end{equation}
If the condensate is highly incompressible (i.e., like for superfluid helium) with a constant density $\rho$, the integral in the square brackets in Eq.~\ref{Eq-AM} reduces to $\rho\Gamma$, where $\Gamma$ is the circulation of the flow along the ring. $\Gamma$ equals the quantum circulation $\kappa$ multiplied by the number of quantized vortices enclosed by the ring. Therefore, if there are no vortices present in the condensate, the angular momentum contained in the ring area in Fig.~\ref{Fig9} should be zero. One can repeat this calculation for all concentric rings in the axially symmetric condensate. As long as the density $\rho$ is constant, the condensate cannot carry angular momentum without quantized vortices in it.
However, for BECs or other superfluids that are highly compressible, the condensate density $\rho$ may evolve with time and can become spatially nonuniform. As a consequence, the integral in Eq.~\ref{Eq-AM} is not necessarily zero, even when the circulation $\Gamma$ is zero. Specifically, for the case that we are considering here, the major part of the yellow ring carries a flow with a tangential velocity in the counterclockwise direction. But due to the opposite phase gradient, the velocity of the yellow ring segment inside the soliton region is clockwise. An integration of the velocity along the complete ring therefore cancels when the circulation $\Gamma=0$. However, when we evaluate the angular momentum, the condensate density is nearly depleted in the soliton region but is finite in the rest part of the ring. As a consequence, the integral in the square brackets in Eq.~\ref{Eq-AM} for evaluating angular momentum becomes finite.

\section {Conclusion}\label{Sec-V}
We have conducted a systematic numerical study on the process of a disk condensate merging with a concentric ring condensate. Our simulation results show that when there exists relative motion between the two condensates, spiral solitons can emerge. These solitons have sharp inner endpoints that can induce rotational motion in the condensate like vortices but with effectively a fraction of a quantum circulation. The existence of these spiral solitons facilitate the transfer of angular momentum between the two condensate regions and allows the merged condensate to carry angular momentum, even in the absence of quantized vortices. Our study also shows that the number of spiral solitons created during condensate merging matches with the initial relative winding number of the two concentric condensates. The underlying mechanism for which the solitons can break up and develop endpoints in the condensate is explained as due to the relative shear flows of the condensates at the interface. This mechanism is general and should be applicable to condensates in 3D as well. Indeed, we have also simulated the merging of a 3D rotating cylindrical condensate with a static cylindrical condensate. A helical soliton sheet appears during the merging whose sharp leading edge can induce flows like a vortex line. In a more general view, it is well known in classical fluids that when there is velocity shear in a single fluid, or when there is a velocity difference across the interface between two fluids, the so-called Kelvin-Helmholtz (KH) instability can occur, which leads to the formation of periodic vortical structures at the interface \cite{Helmholtz-book, Kelvin-book}. In superfluids, KH instability has been observed at the interfaces between two superfluid components, such as at the interface between superfluid $^{3}$He-A and superfluid $^{3}$He-B \cite{Blaauwgeers-2002-PRL, Volovik-2002-JETP}, and in two-component BECs \cite{Takeuchi-2010-PRB, Lundh-2012-PRA}. The mechanism that we have identified should be responsible for the KH instability in a single component superfluid. We shall discuss the details in a future publication.


\begin{acknowledgments}
W. G. acknowledges the support by the National Science Foundation under Grant No. DMR-1507386 and the support from the National High Magnetic Field Laboratory, which is supported by NSF Grant No. DMR-1644779 and the state of Florida. M. T. would like to acknowledge the support by the Japan Society for the Promotion of Science (JSPS) KAKENHI under Grant No. JP17K05548 and JP16H00807.
\end{acknowledgments}

\end{document}